\documentclass[11pt,a4paper]{article}
\usepackage[margin=1in]{geometry}
\usepackage[T1]{fontenc}
\usepackage[utf8]{inputenc}
\usepackage{amsmath,amssymb,amsthm,mathtools,mathrsfs}
\usepackage{bm}
\usepackage{graphicx}
\usepackage{float}
\usepackage{array,tabularx,booktabs}
\usepackage{xcolor}
\usepackage[colorlinks=true,linkcolor=blue!55!black,citecolor=blue!55!black,urlcolor=blue!55!black]{hyperref}
\usepackage{microtype}

\newcommand{\keywords}[1]{\par\medskip\noindent\textbf{Keywords:} #1\par}

\title{\bfseries Charge-Induced Pole Cancellation and Horizon Transitions in Scale-Dependent Gravitational Collapse}
\author{Ghulam Muhammad$^{1}$, Syed Zaheer Abbas$^{1,2}$\thanks{Corresponding author: \href{mailto:zaheer@hu.edu.pk}{zaheer@hu.edu.pk}}, and Muhammad Sajjad$^{1}$\\[0.5em]
\small $^{1}$Department of Mathematics, Hazara University, Mansehra, Pakistan\\
\small $^{2}$Laser Spectroscopy Lab, Centralized Resource Laboratory (CRL), Hazara University, Mansehra 21300, Pakistan}
\date{}

\begin{document}
\maketitle

\begin{abstract}
We construct a charged Oppenheimer--Snyder-like collapse model in scale-dependent gravity by matching a spatially flat FLRW interior to a charged scale-dependent exterior across a timelike thin shell. The electric charge is confined to the stellar surface, preserving interior homogeneity and isotropy. The exterior geometry is supported by a phenomenological Bianchi-consistent effective source, while the shell dynamics follow from the Israel--Maxwell junction conditions. A barotropic surface equation of state closes the shell system, with a charged-dust shell as the minimal realization. For a negative scale-dependent parameter, $\tilde{\omega}<0$, the exterior contains a finite-radius boundary $x_s$ defined by $D(x_s)=0$. Charge separates the solutions into three regimes. For $0\le q^2<x_s$, the lapse develops a negative pole at a curvature singularity, the physical exterior contains one outer horizon, and a representative monotonic collapse crosses this horizon before reaching $x_s$; no future-directed locally outgoing radial null branch emerges from the singular boundary. At $q^2=x_s$, simultaneous zeros of the numerator and denominator cancel the curvature pole, although the prescribed running coupling remains singular. For $q^2>x_s$, the curvature singularity persists with a positive pole and locally outgoing radial null branches exist. Depending on the physical extremality condition $x_e>x_s$, the exterior may contain two simple horizons, one degenerate horizon, or no horizon. These results show that charge qualitatively changes the singular and horizon structure of scale-dependent collapse and provide model-level evidence for horizon shielding in the negative-pole regime.
\end{abstract}


\section{Introduction}
Gravitational collapse is one of the principal mechanisms through which
sufficiently compact matter configurations form black holes. The classical
Oppenheimer--Snyder model provides the simplest exact realization of this
process: a homogeneous and isotropic pressureless FLRW interior is matched
across a timelike stellar surface to a Schwarzschild exterior
\cite{OppenheimerSnyder1939}. Although highly idealized, this construction
remains an important benchmark because it permits the evolution of the stellar
surface, the formation of trapped regions, and the causal relation between the
horizon and the final singular state to be studied analytically.

The singularity theorems establish that geodesic incompleteness can arise under
broad geometric and energy conditions \cite{Penrose1965}. They do not,
however determine whether the resulting singular region is visible to distant
observers. This question motivates the weak cosmic censorship conjecture
according to which singularities generated by physically reasonable
gravitational collapse should be hidden within black hole regions
\cite{Penrose1969,Wald1997}. In concrete collapse models, it is important to
distinguish several logically different statements. The existence of an outer
horizon, the formation of trapped surfaces, the local existence of outgoing
null curves near a singular boundary and the global question of whether any
causal signal reaches future null infinity. Local non-emission or horizon crossing can provide evidence for shielding but neither constitutes by itself
For related treatments of gravitational energy in spherical symmetry and black-hole mechanics, see Refs. \cite{Hayward1996, Poisson2004}.
Classical general relativity is expected to require modification when curvature
approaches the quantum gravity regime. One framework for investigating such
effects is the asymptotic safety program in which Newton's constant and other
gravitational couplings acquire a renormalization scale dependence
\cite{Reuter1998,ReuterSaueressig2012}. Renormalization group improvement of
classical black hole metrics has shown that a running Newton coupling can
substantially modify the short-distance geometry, the number of horizons and
the causal structure \cite{BonannoReuter2000}. A related scale-dependent
approach promotes the gravitational couplings to spacetime dependent
quantities and supplements Einstein's equations with the effective terms
required for consistency. Static black hole solutions with scale-dependent
couplings have been constructed in both neutral and charged settings
\cite{ContrerasKochRioseco2013,KochRioseco2016}. More general analyses have
also emphasized that varying couplings must be accompanied by modified field
equations or conservation relations rather than being introduced as an
unconstrained substitution into a classical metric
\cite{BonannoKofinasZarikas2021}.

An Oppenheimer--Snyder like collapse model in scale-dependent gravity was
recently developed by matching a spatially flat FLRW interior to an improved
static exterior \cite{HassannejadEtAl2025}. In that construction, the sign of
the scale-dependent parameter plays a decisive role. For a negative parameter,
the running gravitational coupling develops a pole at a finite positive
radius $x=x_s$ determined by $D(x_s)=0$. The corresponding exterior geometry
can therefore terminate at a finite radius singular boundary rather than only
at the usual central singularity. The uncharged analysis established the
surface dynamics and examined the event and apparent horizons providing the
direct starting point for the charged model considered here.

The inclusion of electric charge introduces both physical and geometrical
complications. A nonzero radial electric field throughout an exactly
homogeneous and isotropic FLRW region would select a preferred spatial
direction and is therefore incompatible with the assumed interior symmetry.
A natural alternative is to keep the FLRW interior electrically neutral and
place the total charge on the stellar boundary. The boundary must then carry a
Maxwell surface current, while any material surface energy density and
tangential pressure are described independently by a thin-shell stress tensor.
The discontinuity of the electromagnetic field is governed by the Maxwell
junction condition, whereas the discontinuity of the extrinsic curvature is
governed by the Israel equations \cite{Israel1966}. Classical studies of
charged spherical shells have already shown that electric charge can alter the
collapse trajectory, produce turning points and change the relation between
the shell and the Reissner--Nordstr\"om horizons
\cite{deLaCruzIsrael1967,Boulware1973}.

Quantum corrected Oppenheimer--Snyder models have recently provided additional
motivation for studying charged interfaces. A loop quantum cosmology inspired
Oppenheimer--Snyder geometry was constructed by matching a quantum corrected
homogeneous interior to a deformed Schwarzschild exterior
\cite{LewandowskiEtAl2023}. Charged extensions subsequently introduced
electric or magnetic charge on massive timelike shells and investigated
equilibrium, oscillatory and nonsingular configurations
\cite{Mazharimousavi2025EPJC,Mazharimousavi2025Universe}. These models
demonstrate that a charged quantum-corrected boundary can have dynamics that
are qualitatively different from those of a neutral dust surface. They are based on quantum Oppenheimer-Snyder geometries and nonlinear electromagnetic or loop-inspired corrections rather than on the
negative-$\widetilde{\omega}$ finite-radius boundary studied in scale-dependent gravity.

The purpose of the present work is to construct and analyze a charged extension
of the finite radius scale-dependent collapse model characterized by an
exterior lapse function $f_q(x)$. We do not assume that this lapse follows from the standard scale-dependent Einstein–Maxwell equations derived from an action.
Instead, it is supported by a phenomenological, equation-level effective source
construction in which the radial dependence of Newton's coupling generates an
anisotropic polarization tensor. The tensor is chosen so that the complete
equations satisfy the contracted Bianchi identity and reduce to the classical
Einstein–Maxwell system when the running of the coupling is removed.

The neutral FLRW interior is joined to this exterior across a timelike charged
shell. Both Israel equations and the shell conservation law are retained. To
close the material dynamics, we impose a barotropic surface equation of state
$\Pi=w_{\mathrm{s}}\sigma$. The shell energy density is then determined as a
function of its radius and the evolution depends on a finite set of
parameters rather than on an arbitrary function $\sigma(X)$. The minimal
realization is a charged dust shell whose effective potential is $V_q(X)$.

The present study therefore has four principal objectives. First it supplies
a Bianchi-consistent effective source for the charged scale-dependent exterior. Secondly it closes the shell dynamics using the full Israel-Maxwell junction system, shell conservation and a surface equation of state. Third, it derives the corrected extremality condition and classifies the
negative pole, cancellation, two-horizon, extremal and horizonless regions.
Finally, it compares the local outgoing-null behavior on the two sides of the
cancellation curve and identifies the domain in which the numerical collapse
provides evidence for horizon shielding.

The remainder of the paper is organized as follows. We first introduce the
effective exterior equations and derive the charged lapse and polarization
tensor. We then formulate the Israel-Maxwell matching conditions and obtain
the closed shell dynamics. The geometry at $x=x_s$, the exceptional
cancellation condition and the physical extremality criterion are analyzed
next. We subsequently classify the horizon structure and study the local
outgoing-null equations. Numerical examples and the corrected phase diagram
are then presented, followed by a discussion of the scope and limitations of
the resulting horizon-shielding argument.

\section{Geometrical Formulation}\label{geometrical-formulation}

We denote Newton's gravitational constant by $G_0$, the speed of light
by $c$, and the reduced Planck constant by $\hbar$. The radial
coordinate is denoted by $r$, while $M$ represents the mass of the
collapsing object. In natural units, $2M$ corresponds to the
Schwarzschild radius. The symbol $Q$ denotes the electric charge of
the collapsing object, and the dimensionless charge parameter is defined
by $q=Q/(2M)$. A larger value of $q^2$ indicates a greater electric
charge relative to the mass, whereas a smaller value indicates a weaker
electromagnetic influence on the collapse. The proper time measured by
an observer comoving with the collapsing matter is denoted by $\tau$. Throughout this work, we use natural units,
$G_0=c=\hbar=1$. Let $\tilde{\omega}$ be a dimensionless scale-dependent parameter that
characterizes the strength of the quantum-gravity corrections to
Newton's gravitational coupling. In the neutral limit $q\rightarrow0$,
at fixed shell mass, one recovers an uncharged material shell. Recovering the shell-free uncharged matching requires the additional limit $\mu_i\rightarrow0$.

\paragraph*{A Bianchi-consistent Effective Source Construction.}\label{a-bianchi-consistent-effective-field-theory}
We adopt a phenomenological effective field equation in which the prescribed running coupling and a reconstructed polarization tensor support the charged lapse. The construction is Bianchi-consistent but is not assumed to coincide with the standard action-based scale-dependent Einstein–Maxwell formulation. We define the effective field equations by
\begin{equation}
G^{\mu}{}_{\nu}=8\pi G(r)T^{\mathrm{EM}\,\mu}{}_{\nu}+\Theta^{\mu}{}_{\nu}.
\label{eq:eff-field-eq}
\end{equation}
Here, $\Theta^{\mu}{}_{\nu}$ represents vacuum polarization or scale-dependent gravitational effects. In the exterior region, the Maxwell equations remain conserved $(\nabla_{\mu}F^{\mu\nu}=0)$, while the shell provides the distributional electromagnetic source at the stellar boundary.

For the Maxwell region, we choose the electromagnetic potential $A=-\frac{Q}{r}\,dt$. The corresponding electromagnetic field strength is $F=dA=\frac{Q}{r^2}\,dr\wedge dt$. The mixed components of the electromagnetic energy-momentum tensor are
\begin{equation}
T^{\mathrm{EM}\,\mu}{}_{\nu}=\frac{Q^2}{8\pi r^4}\operatorname{diag}(-1,-1,1,1).
\label{eq:TEM}
\end{equation}
No electric volume charge is introduced in the exterior region. For the quantum-polarization tensor, we specify the scale-dependent source by
\begin{equation}
\Theta^{t}{}_{t}=\Theta^{r}{}_{r}=\frac{Q^2-2Mr}{r^3}G'(r),
\label{eq:theta-rad}
\end{equation}
and
\begin{equation}
\Theta^{\theta}{}_{\theta}=\Theta^{\phi}{}_{\phi}=\frac{Q^2-2Mr}{2r^2}G''(r)-\frac{Q^2}{r^3}G'(r).
\label{eq:theta-ang}
\end{equation}
This tensor has three useful properties. First, $\Theta^{t}{}_{t}=\Theta^{r}{}_{r}$, so the Schwarzschild gauge condition $g_{tt}g_{rr}=-1$ remains admissible. Second, the effective tensor vanishes smoothly in the classical limit: as $G'(r)\to 0$ and $G''(r)\to 0$, we have $\Theta^{\mu}{}_{\nu}\to 0$, so the classical Einstein--Maxwell limit is recovered. Third, the tensor satisfies the modified conservation law required by the Bianchi identity.

\paragraph*{Derivation of the Lapse Function.}\label{derivation-of-the-lapse-function}
Consider the static, spherically symmetric exterior line element
$
ds^2=-f(r)\,dt^2+\frac{dr^2}{f(r)}+r^2d\Omega^2.
$
The independent mixed components of the Einstein tensor are
$
G^{t}{}_{t}=G^{r}{}_{r}=\frac{rf'(r)+f(r)-1}{r^2},
$
and
$
G^{\theta}{}_{\theta}=G^{\phi}{}_{\phi}=\frac{f''(r)}{2}+\frac{f'(r)}{r}.
$
The $tt$ component of the effective field equations becomes
$
\frac{rf'(r)+f(r)-1}{r^2}=-\frac{G(r)Q^2}{r^4}+\frac{Q^2-2Mr}{r^3}G'(r).
$
After simplification and integration, we obtain
$
r\bigl(f(r)-1\bigr)=-2MG(r)+\frac{Q^2G(r)}{r}+C.
$
The integration constant $C$ generates an additional Schwarzschild-type term $C/r$, and the boundary condition that the asymptotic mass parameter be $M$ fixes $C=0$, giving
\begin{equation}
f(r)=1-\frac{2MG(r)}{r}+\frac{Q^2G(r)}{r^2}.
\label{eq:fr-lapse}
\end{equation}
The angular field equation is then satisfied by the tangential component of the quantum-polarization tensor given in Eq.~\eqref{eq:theta-ang}.

\paragraph*{Bianchi-Identity Check.}\label{bianchi-identity-check}
The contracted Bianchi identity gives $\nabla_{\mu}G^{\mu}{}_{\nu}=0$. Because the electromagnetic energy-momentum tensor is independently conserved, $\nabla_{\mu}T^{\mathrm{EM}\,\mu}{}_{\nu}=0$, the effective scale-dependent source must satisfy
\begin{equation}
\nabla_{\mu}\Theta^{\mu}{}_{\nu}=-8\pi T^{\mathrm{EM}\,\mu}{}_{\nu}\,\partial_{\mu}G.
\label{FE1}
\end{equation}
For the diagonal source defined above, the only nontrivial component is the radial one:
\begin{equation}
\frac{d\Theta^{r}{}_{r}}{dr}+\frac{2}{r}\left(\Theta^{r}{}_{r}-\Theta^{\theta}{}_{\theta}\right)=\frac{Q^2G'(r)}{r^4}.
\label{eq:bianchi-radial}
\end{equation}
Indeed,
$
-8\pi T^{\mathrm{EM}\,r}{}_{r}G'(r)=-8\pi\left(-\frac{Q^2}{8\pi r^4}\right)G'(r)=\frac{Q^2G'(r)}{r^4},
$
which is precisely the radial component of \eqref{FE1}, demonstrating that the construction is Bianchi-consistent.

\paragraph*{Dimensionless Form.}\label{dimensionless-form}
We introduce the dimensionless quantities
$
L=2M, \qquad x=\frac{r}{L}, \qquad q=\frac{Q}{L}, \qquad d\tilde{s}^2=\frac{ds^2}{L^2},
$
and define the dimensionless running gravitational coupling by
$
g(x)=\frac{G(x)}{G_0}=\frac{x^3}{D(x)},
$
where
\begin{equation}
D(x)=x^3+\tilde{\omega}\left(x+\frac{\gamma}{2}\right).
\label{eq:Dx}
\end{equation}
Equation~\eqref{eq:fr-lapse} then becomes
\begin{equation}
f_q(x)=1-\frac{g(x)}{x}+\frac{q^2g(x)}{x^2}.
\label{eq:fq-inter}
\end{equation}
Substituting $g(x)=x^3/D(x)$ gives
$
f_q(x)=1-\frac{x^2-q^2x}{D(x)}.
$
Equivalently,
\begin{equation}
f_q(x)=\frac{H_q(x)}{D(x)},
\label{eq:fq}
\end{equation}
where
\begin{equation}
H_q(x)=x^3-x^2+\bigl(\tilde{\omega}+q^2\bigr)x+\frac{\tilde{\omega}\gamma}{2}.
\label{eq:Hq}
\end{equation}
This expression, Eq.~\eqref{eq:fq}, reproduces the exterior lapse function used throughout the collapse model and is referenced directly (rather than re-derived) wherever it recurs below.

The dimensionless components of the quantum-polarization tensor are
$\Theta^{t}{}_{t}=\Theta^{r}{}_{r}=\frac{1}{L^2}\frac{q^2-x}{x^3}g_{,x},$
and
$\Theta^{\theta}{}_{\theta}=\Theta^{\phi}{}_{\phi}=\frac{1}{L^2}\left[\frac{q^2-x}{2x^2}g_{,xx}-\frac{q^2}{x^3}g_{,x}\right].$
For the specific running coupling adopted here,
$g_{,x}=\frac{\tilde{\omega}x^2(4x+3\gamma)}{2D(x)^2}.$
All components of the quantum-polarization tensor vanish in the classical limit $\tilde{\omega}\to 0$. The parameter $\tilde{\omega}$ controls the scale dependence of gravity. The two relevant limiting cases are
$q=0 \quad \Longrightarrow \quad f_0(x)=1-\frac{g(x)}{x},$
and $\tilde{\omega}=0 \quad \Longrightarrow \quad f_q(x)=1-\frac{1}{x}+\frac{q^2}{x^2}.$
Thus, the model reduces to uncharged scale-dependent collapse when $q=0$ and to classical Reissner--Nordstr\"om collapse when $\tilde{\omega}=0$.

\section{Closed Charged-Shell Dynamics}\label{closed-charged-shell-dynamics}

We work with the dimensionless variables
$
X=\frac{R}{2M}, \qquad \tilde{\tau}=\frac{\tau}{2M}.
$
An overdot denotes differentiation with respect to $\tilde{\tau}$: $\dot X\equiv dX/d\tilde{\tau}$. The dimensionless surface energy density and surface pressure are defined by
$
\sigma=(2M)\sigma_{\mathrm{phys}}, \qquad \Pi=(2M)\Pi_{\mathrm{phys}}.
$
The surface stress tensor and surface electric current are
$
S^{i}{}_{j}=\operatorname{diag}(-\sigma,\Pi,\Pi), \qquad j^i=\sigma_e u^i.
$

\paragraph*{Maxwell Junction Condition and Charge Conservation.}\label{maxwell-junction-condition-and-charge-conservation}

The Maxwell junction condition across the timelike shell is
\begin{equation}
n_{\mu}\left(F_+^{\mu\nu}-F_-^{\mu\nu}\right)=4\pi j^{\nu}.
\label{eq:maxwell-junction}
\end{equation}
Because the FLRW interior is electrically neutral, $F_-^{\mu\nu}=0$, whereas the exterior electromagnetic field is $F_+=\frac{q}{X^2}\,dX\wedge d\tilde t$. It follows that the surface charge density is
\begin{equation}
\sigma_e(X)=\frac{q}{4\pi X^2}.
\label{eq:sigmae}
\end{equation}
Equivalently, intrinsic charge conservation on the shell gives $D_i j^i=0$, or $\dfrac{d}{d\tilde{\tau}}\left(4\pi X^2\sigma_e\right)=0$. Therefore,
\begin{equation}
q=4\pi X^2\sigma_e
\label{eq:q-const}
\end{equation}
is constant throughout the collapse.

\paragraph*{Israel Junction Equations.}\label{israel-junction-equations}

We choose the unit normal to point from the FLRW interior toward the exterior and consider the ordinary orientation branch. Define
\begin{equation}
\beta_+=\sqrt{\dot X^2+f_q(X)}.
\label{eq:beta-plus}
\end{equation}
For a comoving boundary in a spatially flat FLRW interior, the nonvanishing mixed components of the extrinsic curvature are
$
K^{-\,\theta}{}_{\theta}=K^{-\,\phi}{}_{\phi}=\frac{1}{X}, \qquad K^{-\,\tilde{\tau}}{}_{\tilde{\tau}}=0.
$
The charged exterior metric is
$
ds_+^2=-f_q(X)\,d\tilde t^2+\frac{dX^2}{f_q(X)}+X^2d\Omega^2,
$
with $f_q$ given by Eq.~\eqref{eq:fq}. The corresponding exterior extrinsic-curvature components are
$
K^{+\,\theta}{}_{\theta}=K^{+\,\phi}{}_{\phi}=\frac{\beta_+}{X}, \qquad
K^{+\,\tilde{\tau}}{}_{\tilde{\tau}}=\frac{\ddot X+\tfrac12 f_q'(X)}{\beta_+}.
$
The Israel junction conditions are
\begin{equation}
[K^{i}{}_{j}]-\delta^{i}{}_{j}[K]=-8\pi S^{i}{}_{j}, \qquad [K^{i}{}_{j}]\equiv K^{+\,i}{}_{j}-K^{-\,i}{}_{j}.
\label{eq:israel}
\end{equation}
The angular junction equation gives $4\pi X\sigma=1-\beta_+$, or, equivalently,
\begin{equation}
\beta_+=1-4\pi X\sigma.
\label{eq:beta-sigma}
\end{equation}
The temporal junction equation determines the surface pressure:
\begin{equation}
8\pi\Pi=\frac{\ddot X+\tfrac12 f_q'(X)}{\beta_+}+\frac{\beta_+-1}{X}.
\label{eq:Pi-junction}
\end{equation}
Using Eq.~\eqref{eq:beta-sigma}, this can also be written as
$
\Pi=\frac{1}{8\pi}\frac{\ddot X+\tfrac12 f_q'(X)}{1-4\pi X\sigma}-\frac{\sigma}{2}.
$
Thus, $\Pi$ is determined by the junction equations and is not an independent, unspecified quantity.

\paragraph*{Shell Energy Conservation.}\label{shell-energy-conservation}
The contracted Codazzi identity gives
\begin{equation}
D_iS^{i}{}_{j}=-\left[T^{\mathrm{eff}}_{\mu\nu}n^{\mu}e^{\nu}{}_{j}\right],
\label{eq:codazzi}
\end{equation}
where $T^{\mathrm{eff}}_{\mu\nu}$ includes the exterior electromagnetic and scale-dependent polarization contributions. For $j=\tilde{\tau}$, one obtains
\begin{equation}
\dot\sigma+2\frac{\dot X}{X}(\sigma+\Pi)=-\left[T^{\mathrm{eff}}_{\mu\nu}n^{\mu}u^{\nu}\right].
\label{eq:sigma-dot-flux}
\end{equation}
The right-hand side vanishes in the present model. The comoving FLRW perfect fluid has no radial energy flux, the radial Maxwell field has no Poynting flux in the shell frame, and the effective exterior source satisfies $T^{t}{}_{t}=T^{X}{}_{X}$, which also implies a vanishing $u^{\mu}n^{\nu}$ energy flux. Therefore,
\begin{equation}
\dot\sigma+2\frac{\dot X}{X}(\sigma+\Pi)=0.
\label{eq:sigma-conservation}
\end{equation}
For $\dot X\neq 0$, this relation becomes $\dfrac{d\sigma}{dX}+\dfrac{2}{X}(\sigma+\Pi)=0$, equivalently, the shell first law $\dfrac{d}{d\tilde{\tau}}\left(4\pi X^2\sigma\right)+\Pi\dfrac{d}{d\tilde{\tau}}\left(4\pi X^2\right)=0$.

The Israel junction equations and the conservation equation must be supplemented by a phenomenological matter relation to close the shell dynamics.

\paragraph*{Barotropic Equation-of-State Closure.}\label{barotropic-equation-of-state-closure}
A simple physical closure is the barotropic surface equation of state $\Pi=w_s\sigma$, where $w_s$ is a constant surface equation-of-state parameter. The conservation equation, Eq.~\eqref{eq:sigma-conservation}, then becomes $\dfrac{d\sigma}{dX}+\dfrac{2(1+w_s)}{X}\sigma=0$. Integrating from the initial radius $X_i$ gives
\begin{equation}
\sigma(X)=\sigma_i\left(\frac{X_i}{X}\right)^{2(1+w_s)}.
\label{eq:sigma-X}
\end{equation}
Define the dimensionless material energy of the shell by $\mu(X)=4\pi X^2\sigma(X)$. It then follows that
\begin{equation}
\mu(X)=\mu_i\left(\frac{X_i}{X}\right)^{2w_s}, \qquad \mu_i=4\pi X_i^2\sigma_i.
\label{eq:mu-X}
\end{equation}
The angular junction equation, Eq.~\eqref{eq:beta-sigma}, becomes
\begin{equation}
\sqrt{\dot X^2+f_q(X)}=1-\frac{\mu(X)}{X}.
\label{eq:angular-mu}
\end{equation}
This relation must be retained to select the correct orientation branch; in particular, it requires $1-\mu(X)/X\geq 0$ and $\mu(X)\geq 0$. Squaring Eq.~\eqref{eq:angular-mu} gives the closed equation of motion
\begin{equation}
\dot X^2=V_q(X)=\left[1-\frac{\mu_i}{X}\left(\frac{X_i}{X}\right)^{2w_s}\right]^2-f_q(X).
\label{eq:Vq-general}
\end{equation}
For the collapsing branch, $\dot X=-\sqrt{V_q(X)}$. The shell dynamics are therefore specified by the finite parameter set $\left(\tilde{\omega},\gamma,q,w_s,\mu_i,X_i\right)$.

\paragraph*{Second-Order Equation and Consistency Analysis.}\label{second-order-equation-and-consistency-check}
Differentiating $\dot X^2=V_q(X)$ with respect to $\tilde{\tau}$ gives $\ddot X=\tfrac12 V_q'(X)$ for $\dot X\neq0$. Since $\mu'(X)=-\dfrac{2w_s}{X}\mu(X)$, we obtain
$
\ddot X=(1+2w_s)\frac{\mu(X)}{X^2}\left(1-\frac{\mu(X)}{X}\right)-\frac{1}{2}f_q'(X).
$
Taking $1-\mu(X)/X>0$, substitution into the pressure junction equation, Eq.~\eqref{eq:Pi-junction}, gives
$
8\pi\Pi
=\frac{\ddot X+\tfrac12 f_q'(X)}{1-\mu(X)/X}-\frac{\mu(X)}{X^2}
=(1+2w_s)\frac{\mu(X)}{X^2}-\frac{\mu(X)}{X^2}
=2w_s\frac{\mu(X)}{X^2}.
$
Since $\sigma(X)=\mu(X)/(4\pi X^2)$, the preceding equation gives $\Pi=w_s\sigma$. This confirms that the angular junction equation, the temporal junction equation, the conservation law, and the barotropic equation of state are mutually consistent.

\paragraph*{A Charged Dust Shell: Minimal Model.}\label{minimal-model-a-charged-dust-shell}
The simplest physically interpretable choice is $w_s=0$, i.e.\ $\Pi=0$. The surface energy density then becomes $\sigma(X)=\sigma_i(X_i/X)^2$, while the shell material energy is constant:
\begin{equation}
\mu(X)=\mu_0=\text{constant}.
\label{eq:mu0-const}
\end{equation}
Consequently, $\sigma(X)=\mu_0/(4\pi X^2)$ and $\Pi=0$. The closed collapse equation is
\begin{equation}
\dot X^2=\left(1-\frac{\mu_0}{X}\right)^2-f_q(X),
\label{eq:Vq-dust}
\end{equation}
with collapsing branch $\dot X=-\sqrt{(1-\mu_0/X)^2-f_q(X)}$ and acceleration
$
\ddot X=\frac{\mu_0}{X^2}\left(1-\frac{\mu_0}{X}\right)-\frac{1}{2}f_q'(X).
$
The dynamically allowed region must satisfy $(1-\mu_0/X)^2-f_q(X)\geq 0$, together with the unsquared branch condition $1-\mu_0/X\geq 0$.

\paragraph*{Electromagnetic Force on the Shell.}\label{electromagnetic-force-on-the-shell}

Although the electromagnetic field contributes no term to the intrinsic shell-energy conservation equation, it exerts a normal force on the charged shell. For $E_-=0$ and $E_+=Q/R^2$, the electric field acting on an ideal infinitesimally thin charged layer is the average of the fields on the two sides, $E=(E_++E_-)/2=Q/(2R^2)$, so the resulting outward electromagnetic force per unit area is $f_{\mathrm{EM}}=\sigma_e E=Q^2/(8\pi R^4)$.

This force acts normal to the shell and does not represent an energy flux along the shell. It is already encoded in the charge-dependent exterior lapse $f_q(X)$ and in the normal Israel equation. It must therefore not be added separately to $V_q(X)$, because doing so would double-count the electromagnetic interaction.

The normal stress-balance identity provides an additional consistency check:
\begin{equation}
S^{ij}\bar K_{ij}=\left[T^{\mathrm{eff}}_{\mu\nu}n^{\mu}n^{\nu}\right],
\label{eq:normal-stress}
\end{equation}
where $\bar K_{ij}=\tfrac12(K^+_{ij}+K^-_{ij})$ is the average extrinsic curvature across the shell. The electromagnetic radial stress and the scale-dependent polarization stress both contribute to the right-hand side.

\paragraph*{Initial-Data Relation.}\label{initial-data-relation}

The initial shell density is not independent of both the initial radius $X_i$ and the initial velocity $\dot X_i$. At the initial surface, $\sqrt{\dot X_i^2+f_q(X_i)}=1-\mu_i/X_i$, so that
\begin{equation}
\mu_i=X_i\left[1-\sqrt{\dot X_i^2+f_q(X_i)}\right].
\label{eq:mui-init}
\end{equation}
Equivalently, $\sigma_i=\left[1-\sqrt{\dot X_i^2+f_q(X_i)}\right]/(4\pi X_i)$. One may therefore prescribe either $(X_i,\dot X_i)$ and determine $\mu_i$, or prescribe $(X_i,\mu_i)$ and determine $\dot X_i$. Prescribing all three quantities independently would overdetermine the shell.

\paragraph*{Neutral and Current-Only Limits.}\label{neutral-and-current-only-limits}

For a massive charged shell, the shell-free smooth matching limit requires $q\to 0$ and $\mu_i\to 0$ jointly. Taking only $q\to 0$ while keeping $\mu_i$ fixed is the thin-shell model, yielding an uncharged material shell and therefore not recovering the smooth uncharged collapse model. A fixed charge-to-mass ratio may be introduced through $\zeta=|q|/\mu_0$, i.e.\ $\mu_0=|q|/\zeta$, so that $q\to 0 \Rightarrow \mu_0\to 0$.

A separate and internally consistent Maxwell-current-only model is obtained by setting $\sigma=\Pi=0$, $\sigma_e=q/(4\pi X^2)$. The angular junction equation then gives $\dot X^2=1-f_q(X)$. This configuration contains a charged boundary current but no gravitational thin shell, and should therefore not be described as a massive charged shell.

\paragraph*{Interior Geometry.}\label{interior-geometry}

The interior spacetime metric is
\begin{equation}
ds_-^2=-d\tilde{\tau}^2+a^2(\tilde{\tau})\left(d\chi^2+\chi^2d\Omega^2\right).
\label{eq:interior-metric}
\end{equation}
The stellar surface is located at $\chi=\chi_b$, with areal radius $X(\tilde{\tau})=a(\tilde{\tau})\chi_b$. The Hubble parameter is defined by
\begin{equation}
H(\tilde{\tau})=\frac{\dot a}{a}=\frac{\dot X}{X}.
\label{eq:hubble}
\end{equation}
Here, $a(\tilde{\tau})$ is the scale factor, $\dot a$ is its derivative with respect to proper time, and $H(\tilde{\tau})$ measures the contraction rate of the stellar interior.

The interior matter is a perfect fluid, $T^{\mu}{}_{\nu}=\operatorname{diag}(-\rho,p,p,p)$. The improved Friedmann equation is
\begin{equation}
H^2=\frac{8\pi G(X)}{3}\rho.
\label{eq:friedmann}
\end{equation}
Therefore,
\begin{equation}
\rho(X)=\frac{3D(X)}{8\pi X^5}\dot X^2.
\label{eq:rho-X}
\end{equation}
The conservation equation is
\begin{equation}
\dot\rho+3H(\rho+p)=-\rho\frac{\dot G}{G}.
\label{eq:rho-conservation}
\end{equation}
The additional term on the right-hand side appears because the gravitational coupling varies with scale. Thus,
\begin{equation}
p(X)=-\rho(X)-\frac{X}{3}\left[\rho'(X)+\rho(X)\frac{G'(X)}{G(X)}\right].
\label{eq:pressure-X}
\end{equation}

\section{Geometry at $x=x_s$}\label{geometry-at-xs}

Recall that $D(x)$ is given by Eq.~\eqref{eq:Dx}, with $\tilde{\omega}<0$ and $\gamma>0$, and let $x_s>0$ be the unique positive root satisfying
\begin{equation}
D(x_s)=0.
\label{eq:xs-def}
\end{equation}
The lapse function $f_q(x)=H_q(x)/D(x)$ was already given in Eq.~\eqref{eq:fq}, with $H_q(x)$ as in Eq.~\eqref{eq:Hq}. Since
$
H_q(x)=D(x)-x^2+q^2x,
$
evaluation at $x=x_s$ gives, using Eq.~\eqref{eq:xs-def},
$
H_q(x_s)=x_s\left(q^2-x_s\right).
$
Moreover,
\begin{equation}
D'(x_s)=3x_s^2+\tilde{\omega}>0.
\label{eq:Dprime-xs}
\end{equation}
Writing $\epsilon=x-x_s$, the denominator behaves as $D(x)=D_s'\epsilon+O(\epsilon^2)$, with $D_s'\equiv D'(x_s)>0$. Therefore, when $q^2\neq x_s$,
\begin{equation}
f_q(x)=\frac{x_s(q^2-x_s)}{D_s'(x-x_s)}+O(1).
\label{eq:fq-near-xs}
\end{equation}
Thus,
$
q^2<x_s \quad \Longrightarrow \quad f_q(x)\longrightarrow -\infty,
$
and
$
q^2>x_s \quad \Longrightarrow \quad f_q(x)\longrightarrow +\infty.
$
Both cases contain a pole. For example, the Kretschmann scalar of a metric $ds^2=-f(x)\,dt^2+dx^2/f(x)+x^2d\Omega^2$ contains a contribution proportional to $[f''(x)]^2$. Since
$
f_q''(x)\sim\frac{2x_s(q^2-x_s)}{D_s'(x-x_s)^3},
$
we obtain
\begin{equation}
K\propto\frac{1}{(x-x_s)^6}.
\label{eq:kretschmann}
\end{equation}
Consequently, $q^2\neq x_s \Rightarrow x=x_s$ is a curvature singularity. Charge values satisfying $q^2>x_s$ reverse the sign of the pole but do not eliminate the singularity.

\section{The Exceptional Cancellation $q^2=x_s$}\label{exceptional-cancellation}

When $q^2=x_s$, both the numerator and denominator vanish: $H_q(x_s)=D(x_s)=0$. Using l'H\^opital's rule,
\begin{equation}
\lim_{x\to x_s}f_q(x)=\frac{H_q'(x_s)}{D'(x_s)}.
\label{eq:lhopital}
\end{equation}
Because $H_q'(x)=D'(x)+q^2-2x$, the tuned value $q^2=x_s$ gives $H_q'(x_s)=D_s'-x_s$. Hence,
\begin{equation}
f_q(x_s)=1-\frac{x_s}{D_s'}.
\label{eq:fq-at-xs}
\end{equation}
The common factor $x-x_s$ cancels between $H_q$ and $D$, and the quotient admits an analytic extension through $x_s$. Therefore, $q^2=x_s$ removes the curvature pole at $x_s$. This is the only charge value for which the finite-radius curvature singularity disappears.

The running gravitational coupling $G(x)=x^3/D(x)$ still diverges at $x_s$. Thus, $x_s$ is geometrically regular after the cancellation, but the chosen running-coupling parametrization remains singular there.

\section{Physical Extremalities}\label{physical-extremalities}
Horizons away from $D=0$ satisfy $H_q(x_h)=0$. A degenerate horizon at $x=x_e$ satisfies
\begin{equation}
H_q(x_e)=0, \qquad H_q'(x_e)=0.
\label{eq:degenerate-def}
\end{equation}
The derivative condition gives
\begin{equation}
q_{\mathrm{ext}}^2=-3x_e^2+2x_e-\tilde{\omega}.
\label{eq:qext}
\end{equation}
Substitution into $H_q(x_e)=0$ gives the equation
\begin{equation}
2x_e^3-x_e^2-\frac{\tilde{\omega}\gamma}{2}=0.
\label{eq:extremal-cubic}
\end{equation}
However, not every positive solution of this equation represents a physical extremal horizon. The exterior spacetime considered here is restricted to $x>x_s$. Therefore, a physically relevant extremal horizon must satisfy $x_e>x_s$ and $q_{\mathrm{ext}}^2\geq 0$. An algebraic double root with $x_e<x_s$ lies beyond the singular boundary and must not be used to classify the exterior horizon structure.

\section{Existence and Classification of Horizons}\label{existence-and-classification-of-horizons}

Define $F(x)=2x^3-x^2-\tilde{\omega}\gamma/2$, so the extremal radius is a root of $F(x_e)=0$. At $x=x_s$, using $D(x_s)=0$, we find
\begin{equation}
F(x_s)=x_s\left[D'(x_s)-x_s\right].
\label{eq:F-at-xs}
\end{equation}
Thus, a physical extremal root $x_e>x_s$ exists when $D'(x_s)<x_s$. The limiting value is determined by $D'(x_s)=x_s$. Using $\tilde{\omega}=-x_s^3/(x_s+\gamma/2)$, this condition becomes
\begin{equation}
4x_s^2+(3\gamma-2)x_s-\gamma=0.
\label{eq:xc-condition}
\end{equation}
Define
\begin{equation}
x_c=\frac{2-3\gamma+\sqrt{9\gamma^2+4\gamma+4}}{8}.
\label{eq:xc-def}
\end{equation}
Then $D'(x_s)<x_s \Leftrightarrow x_s<x_c$. The corresponding critical scale-dependent parameter is
\begin{equation}
\tilde{\omega}_c=-\frac{x_c^3}{x_c+\gamma/2}.
\label{eq:omegac-def}
\end{equation}
Therefore, $x_s<x_c$ implies a physical extremal curve exists at $x_e>x_s$, whereas $x_s>x_c$ implies no extremal horizon exists in the physical region $x>x_s$. For the value used throughout this manuscript, $\gamma=1$,
$
x_c\simeq 0.390388, \qquad \tilde{\omega}_c\simeq -0.066821.
$
Thus, for $\gamma=1$, the extremal curve is physically relevant only for approximately $-0.066821<\tilde{\omega}<0$. If the phase diagram extends to $\tilde{\omega}=-0.08$, the portion of the algebraic extremal curve below $\tilde{\omega}_c$ must not be interpreted as an exterior degenerate horizon.

\paragraph*{Negative-Pole Region: $0\leq q^2<x_s$.}\label{negative-pole-sector-0leq-q2x_s}

For $0\leq q^2<x_s$, we have $H_q(x_s)<0$ and $H_q(x)\to +\infty$ as $x\to\infty$. Continuity therefore guarantees an outer root satisfying $x_+>x_s$. In the physical region $x>x_s$, there is one outer horizon. The singularity has a negative pole, $f_q(x)\sim -A_q/(x-x_s)$, where
\begin{equation}
A_q=\frac{x_s(x_s-q^2)}{D_s'}>0.
\label{eq:Aq}
\end{equation}
This is the region covered by the manuscript's local non-visibility argument.

\paragraph*{Cancellation Surface: $q^2=x_s$.}\label{cancellation-surface-q2x_s}

At $q^2=x_s$, the finite-radius curvature singularity is removed. The horizon structure depends on the relative magnitudes of $x_s$ and $x_c$.

\paragraph*{Case $x_s<x_c$.}\label{case-x_sx_c}
Here, $D'(x_s)-x_s<0$, and consequently $f_q(x_s)<0$. The lapse subsequently crosses zero at an outer radius $x_+>x_s$. Thus, the regular cancellation surface lies inside an outer horizon.

\paragraph*{Case $x_s=x_c$.}\label{case-x_sx_c-1}
Here, $D'(x_s)=x_s$, so $f_q(x_s)=0$. After cancellation of the common factor in $H_q/D$, the surface $x=x_s$ becomes a regular critical horizon: $q^2=x_s=x_c$ gives a regular horizon-cancellation coincidence.

\paragraph*{Case $x_s>x_c$.}\label{case-x_sx_c-2}
Here, $D'(x_s)-x_s>0$, and no horizon occurs in the physical region $x>x_s$. Nevertheless, $x_s$ itself is not a curvature singularity because the pole has been canceled.

\paragraph*{Positive-Pole Region: $q^2>x_s$.}\label{positive-pole-sector-q2x_s}

For $q^2>x_s$, the finite-radius curvature singularity remains present, and $f_q(x)\to +\infty$ as $x\to x_s^{+}$. The sign conditions at $x_s$ and infinity are both positive: $H_q(x_s)>0$, $H_q(\infty)>0$. Therefore, the existence of horizons is no longer guaranteed by continuity; the polynomial must dip below zero between $x_s$ and infinity.

\paragraph*{Case $x_s<x_c$.}\label{case-x_sx_c-3}
A physical extremal charge exists. Let $x_e>x_s$ be the unique physical solution of Eq.~\eqref{eq:extremal-cubic}, and define
\begin{equation}
q_H^2=-3x_e^2+2x_e-\tilde{\omega}.
\label{eq:qH}
\end{equation}
The notation $q_H$ emphasizes that this quantity is the physical horizon-loss threshold rather than an arbitrary algebraic extremal root.

The classification is then as follows. The interval $x_s<q^2<q_H^2$ contains two simple horizons, $x_s<x_-<x_+$. At $q^2=q_H^2$, the horizons merge into one degenerate horizon, $x_-=x_+=x_e$. For $q^2>q_H^2$, there is no horizon in the physical region $x>x_s$.

\paragraph*{Case $x_s\geq x_c$.}\label{case-x_sgeq-x_c}
No physical extremal root exists outside $x_s$. Consequently, $q^2>x_s$ implies no horizon exists in the physical exterior. Thus, a statement involving $q_{\mathrm{ext}}^2$ is meaningful only after imposing the condition $x_e>x_s$.

\section{Outgoing-Null-Ray Behavior}\label{outgoing-null-ray-behavior}

In ingoing Eddington--Finkelstein coordinates, outgoing radial null rays satisfy
\begin{equation}
\frac{dx}{dv}=\frac{f_q(x)}{2}.
\label{eq:null-eq}
\end{equation}
Near $x=x_s$, define $\epsilon=x-x_s$ and
\begin{equation}
C_q=\frac{x_s(q^2-x_s)}{D_s'}.
\label{eq:Cq}
\end{equation}
Then
$
\frac{d\epsilon}{dv}\sim\frac{C_q}{2\epsilon}.
$
Integration gives $\epsilon^2=C_q(v-v_s)$.

\paragraph*{Case $q^2<x_s$.}\label{case-q2x_s}

Here, $C_q<0$. Writing $A_q=-C_q>0$, we obtain
\begin{equation}
\epsilon^2=A_q(v_s-v).
\label{eq:eps-neg}
\end{equation}
There is no real outgoing branch for $v>v_s$. Therefore, $q^2<x_s$ implies the singularity is locally non-emitting.

\paragraph*{Case $q^2>x_s$.}\label{case-q2x_s-1}

Here, $C_q>0$, and $\epsilon^2=C_q(v-v_s)$. A real outgoing branch exists for $v>v_s$. Therefore, $q^2>x_s$ implies outgoing null rays emerge locally from $x_s$.

For $0\le q^2<x_s$, the singular boundary is locally non-emitting. If the shell remains on the monotonic collapsing branch and crosses the outer exterior horizon before reaching $x_s$, the model provides strong evidence for horizon shielding. The parameter classification is summarized in Table~\ref{tab:corrected-parameter-classification}.

\begin{table*}[htbp]
\centering
\caption{Classification of the charged scale-dependent geometries.}
\label{tab:corrected-parameter-classification}
\renewcommand{\arraystretch}{1.35}
\setlength{\tabcolsep}{5pt}
\begin{tabular}{p{0.18\textwidth} p{0.19\textwidth} p{0.20\textwidth} p{0.18\textwidth} p{0.20\textwidth}}
\hline
\textbf{Regime} & \textbf{Conditions} & \textbf{Behavior at $x_s$} & \textbf{Horizons in $x>x_s$} & \textbf{Interpretation} \\
\hline
Neutral scale-dependent & $q=0$, $\widetilde{\omega}<0$ & Negative-pole curvature singularity & One outer horizon & Locally non-emitting and horizon-shielded \\
Negative-pole charged & $0<q^{2}<x_s$ & Curvature singularity, $f_q(x)\rightarrow-\infty$ & One outer horizon, $x_{+}>x_s$ & Region supported by the shielding argument \\
Cancellation surface & $q^{2}=x_s$ & Curvature pole removed & Depends on the ratio $x_s/x_c$ & No finite-radius curvature singularity \\
Positive-pole, two-horizon & $x_s<q^{2}<q_H^{2}$, with $x_s<x_c$ & Curvature singularity, $f_q(x)\rightarrow+\infty$ & Two horizons, $x_s<x_{-}<x_{+}$ & Locally emitting, but shielded by the outer horizon at the static-geometric level \\
Physical extremal & $q^{2}=q_H^{2}$, with $x_e>x_s$ & Curvature singularity & One degenerate horizon, $x_{-}=x_{+}=x_e$ & Requires a separate extremal-collapse analysis \\
Horizonless positive-pole & $q^{2}>q_H^{2}$, or $q^{2}>x_s$ when $x_s\geq x_c$ & Curvature singularity & No horizon in the physical exterior & Candidate locally visible singularity \\
\hline
\end{tabular}
\end{table*}

\section{Results and Discussion}\label{results-and-discussion}

The figures clarify the phase structure, shell dynamics, and local causal behavior of the charged scale-dependent collapse model. They distinguish three qualitatively different regimes: the negative-pole region $q^{2}<x_s$, the cancellation curve $q^{2}=x_s$, and the positive-pole region $q^{2}>x_s$. This distinction is essential because the sign of the lapse divergence at $x=x_s$ determines both the horizon structure and the local behavior of outgoing radial null curves.

\paragraph*{Phase-Change Diagram.}\label{phase-change-diagram}

\begin{figure}[htbp]
    \centering
    \includegraphics[width=0.7\linewidth]{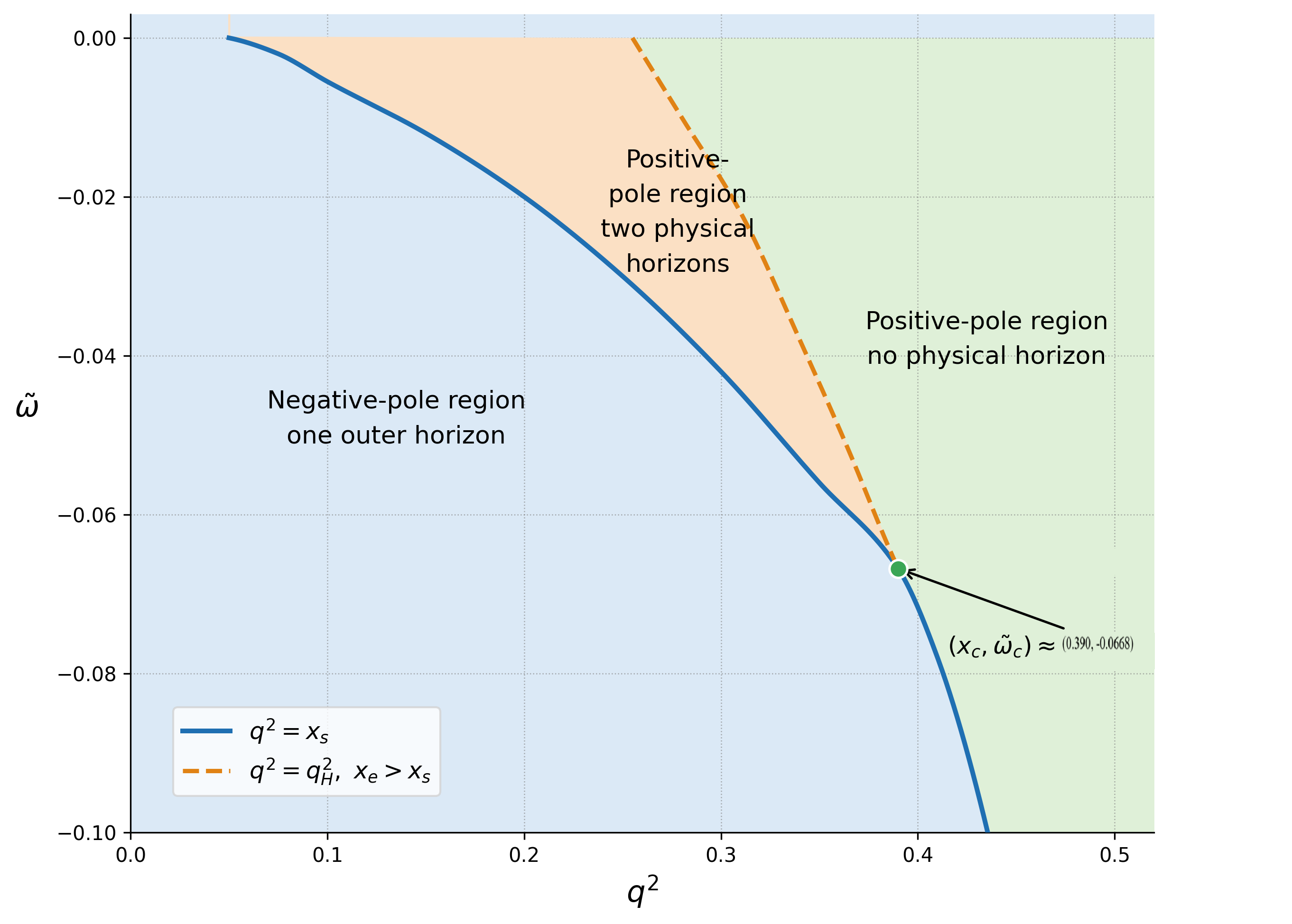}
    \caption{Phase diagram for the $(q^{2},\widetilde{\omega})$ plane for $\gamma=1$.}
    \label{fig:1}
\end{figure}

Figure~\ref{fig:1} presents the phase structure of the scale-dependent exterior in the $(q^{2},\widetilde{\omega})$ plane for $\gamma=1$. The solid curve $q^{2}=x_s$ separates the negative-pole and positive-pole regions associated with the finite-radius boundary $x=x_s$, where $x_s$ is the unique positive root of Eq.~\eqref{eq:xs-def}. Below the solid curve, $q^{2}<x_s$, the numerator of the lapse satisfies $H_q(x_s)=x_s(q^{2}-x_s)<0$. Since $D'(x_s)>0$, the lapse behaves near the singular boundary as in Eq.~\eqref{eq:fq-near-xs}, with $A_q$ given by Eq.~\eqref{eq:Aq}, so $f_q(x)\to-\infty$ as $x\to x_s^{+}$.

In this region, continuity of $H_q(x)$, together with $H_q(x_s)<0$ and $H_q(x)\to+\infty$ as $x\to\infty$, guarantees the existence of an outer horizon satisfying $x_+>x_s$. The root classification shows that the physical exterior contains one outer horizon in this region.

The solid curve itself corresponds to the tuned condition $q^{2}=x_s$. At this value, $H_q(x_s)=D(x_s)=0$, so the zero of $H_q(x)$ coincides with the zero of $D(x)$, and the common factor in $f_q(x)=H_q(x)/D(x)$ cancels; the extended lapse has the finite value given by Eq.~\eqref{eq:fq-at-xs}. Consequently, the finite-radius curvature singularity is absent on the cancellation curve. The cancellation curve should be interpreted as a geometrically exceptional boundary rather than as an ordinary singular region. Nevertheless, the running gravitational coupling $G(x)=x^{3}/D(x)$ remains divergent at $x_s$. Thus, the metric may admit a regular extension through the cancellation surface even though the chosen scale-dependent coupling parametrization remains singular there.

Above the solid curve, $q^{2}>x_s$, the finite-radius curvature singularity persists, but the sign of the lapse divergence changes: $f_q(x)\simeq C_q/(x-x_s)$ with $C_q$ as in Eq.~\eqref{eq:Cq}, so $f_q(x)\to+\infty$ as $x\to x_s^{+}$.

The dashed curve in Figure~\ref{fig:1} denotes the physical extremal threshold $q^{2}=q_H^{2}$, plotted only when the degenerate horizon radius satisfies the physical condition $x_e>x_s$. The extremal radius is determined by Eq.~\eqref{eq:extremal-cubic}, and the corresponding horizon-loss charge is given by Eq.~\eqref{eq:qH}. The region between the cancellation and extremal curves, $x_s<q^{2}<q_H^{2}$, contains two physical horizons, $x_s<x_-<x_+$. At the dashed curve, these horizons merge into a single degenerate horizon, $x_-=x_+=x_e$. For $q^{2}>q_H^{2}$, the exterior contains no horizon in the physical region $x>x_s$.

For $\gamma=1$, the cancellation and physical extremal curves meet at approximately $x_c\simeq0.390388$, $\widetilde{\omega}_c\simeq-0.066821$. Thus, the physical extremal curve is relevant only for $-0.066821<\widetilde{\omega}<0$. For $\widetilde{\omega}<\widetilde{\omega}_c$, no physical extremal radius exists outside $x_s$, and positive-pole configurations in this part of parameter space are therefore horizonless. The phase diagram consequently shows that electric charge does more than shift the position of the outer horizon: it can change the number of horizons, produce a degenerate configuration, or eliminate the horizons altogether, thereby yielding a horizonless finite-radius curvature singularity.

\paragraph*{Evolution of the Closed Charged-Dust Shell.}\label{evolution-of-the-closed-charged-dust-shell}

\begin{figure}[H]
    \centering
    \includegraphics[width=0.7\linewidth]{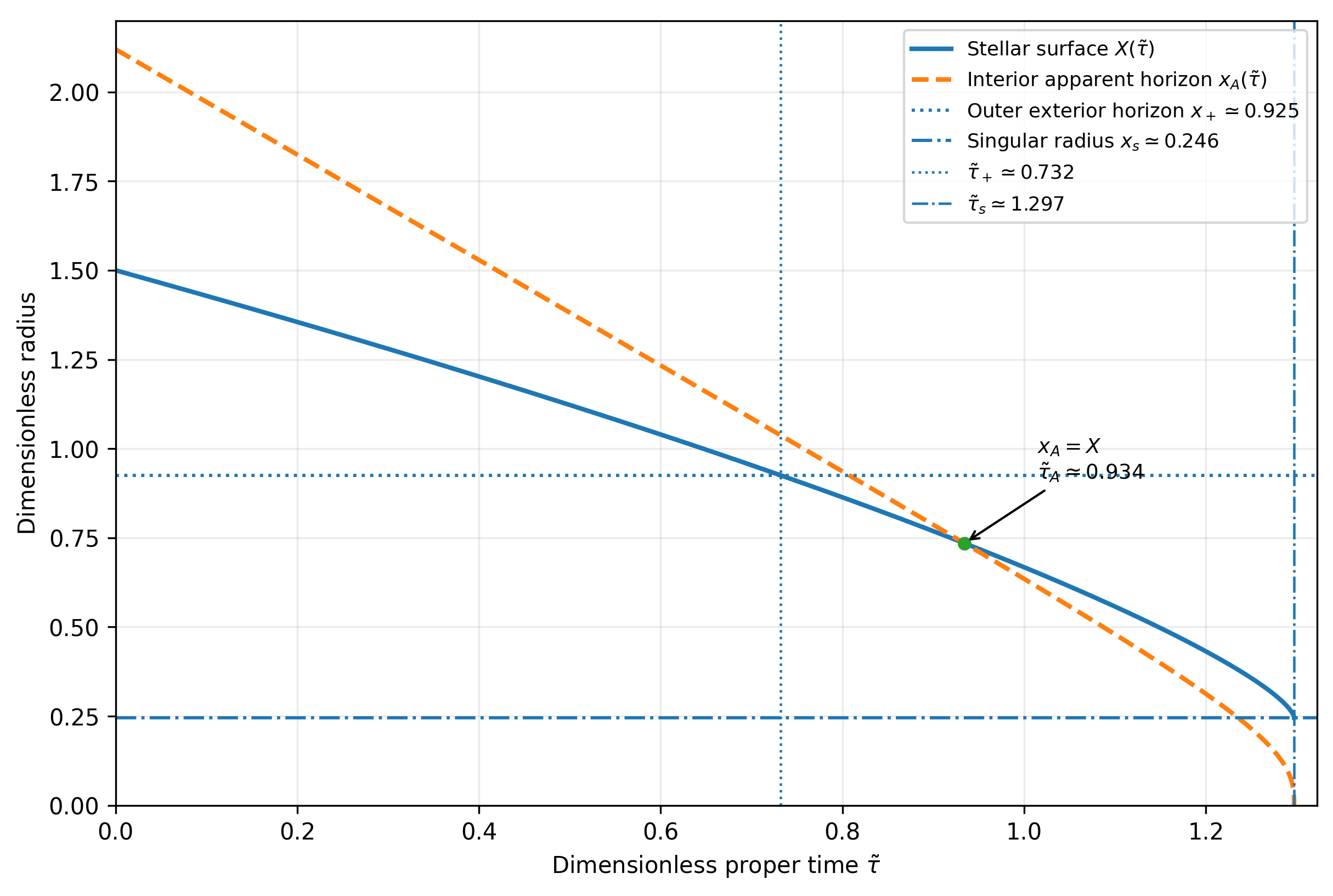}
    \caption{Evolution of the closed charged-dust-shell model for $\gamma=1$, $\widetilde{\omega}=-0.02$, $q^{2}=0.10$, $\mu_0=0.10$, and $X_i=1.50$. The shell evolution is governed by the effective potential in Eq.~\eqref{eq:Vq-dust}.}
    \label{fig:2}
\end{figure}

Figure~\ref{fig:2} displays the evolution of the closed charged-dust-shell model for $\gamma=1$, $\widetilde{\omega}=-0.02$, $q^{2}=0.10$, $\mu_0=0.10$, $X_i=1.50$. For these parameters, $x_s\simeq0.246$, and therefore $q^{2}=0.10<x_s$: the configuration lies in the negative-pole region identified in Figure~\ref{fig:1}. The shell motion is determined by the closed effective potential of Eq.~\eqref{eq:Vq-dust}, which follows from the charged-dust-shell equation of state $\Pi=0$ and the conservation relation $\mu(X)=\mu_0$.

The collapsing branch satisfies $\dot{X}=-\sqrt{V_q(X)}$. The stellar surface decreases monotonically from $X_i=1.50$. Since $V_q(X)$ remains positive throughout the interval $x_s<X\leq X_i$, the chosen initial data produce continuous collapse without a turning point or bounce. The surface crosses the outer exterior horizon $x_+\simeq0.925$ at $\widetilde{\tau}_+\simeq0.732$, and subsequently reaches the finite-radius singularity $x_s\simeq0.246$ at $\widetilde{\tau}_s\simeq1.297$. Thus, $\widetilde{\tau}_s-\widetilde{\tau}_+\simeq0.565>0$: the shell crosses the outer zero of the exterior lapse before reaching the singular boundary, consistent with the analytical ordering $x_s<x_+$, $\widetilde{\tau}_s>\widetilde{\tau}_+$.

The interior apparent-horizon radius is
$
x_A(X) = \frac{X}{\sqrt{V_q(X)}}.
$
At the beginning of the evolution, the apparent horizon lies outside the stellar surface. It decreases as the collapse proceeds and intersects the stellar boundary at approximately $X\simeq0.735$, $\widetilde{\tau}\simeq0.934$, where $x_A=X$, equivalently $V_q(X)=1$. After this time, $x_A<X$, and the stellar surface lies inside the trapped portion of the FLRW interior.

The fact that the stellar surface crosses the exterior horizon before the interior apparent horizon reaches it is not contradictory. The exterior horizon is a Killing horizon of the static exterior geometry. By contrast, the apparent horizon is defined locally by the instantaneous expansion of the null congruences and is generally slicing dependent; these two surfaces need not intersect the shell at the same proper time during dynamical collapse. Near the singular radius, the negative divergence of $f_q(X)$ implies $V_q(X)\to+\infty$ as $X\to x_s^{+}$, so $x_A(X)\to 0$. Accordingly, the trapped region expands through the FLRW interior as the singular state is approached.

\paragraph*{Outgoing Null Curves in the Negative-Pole Region.}\label{outgoing-negative-pole}

\begin{figure}[h]
    \centering
    \includegraphics[width=0.7\linewidth]{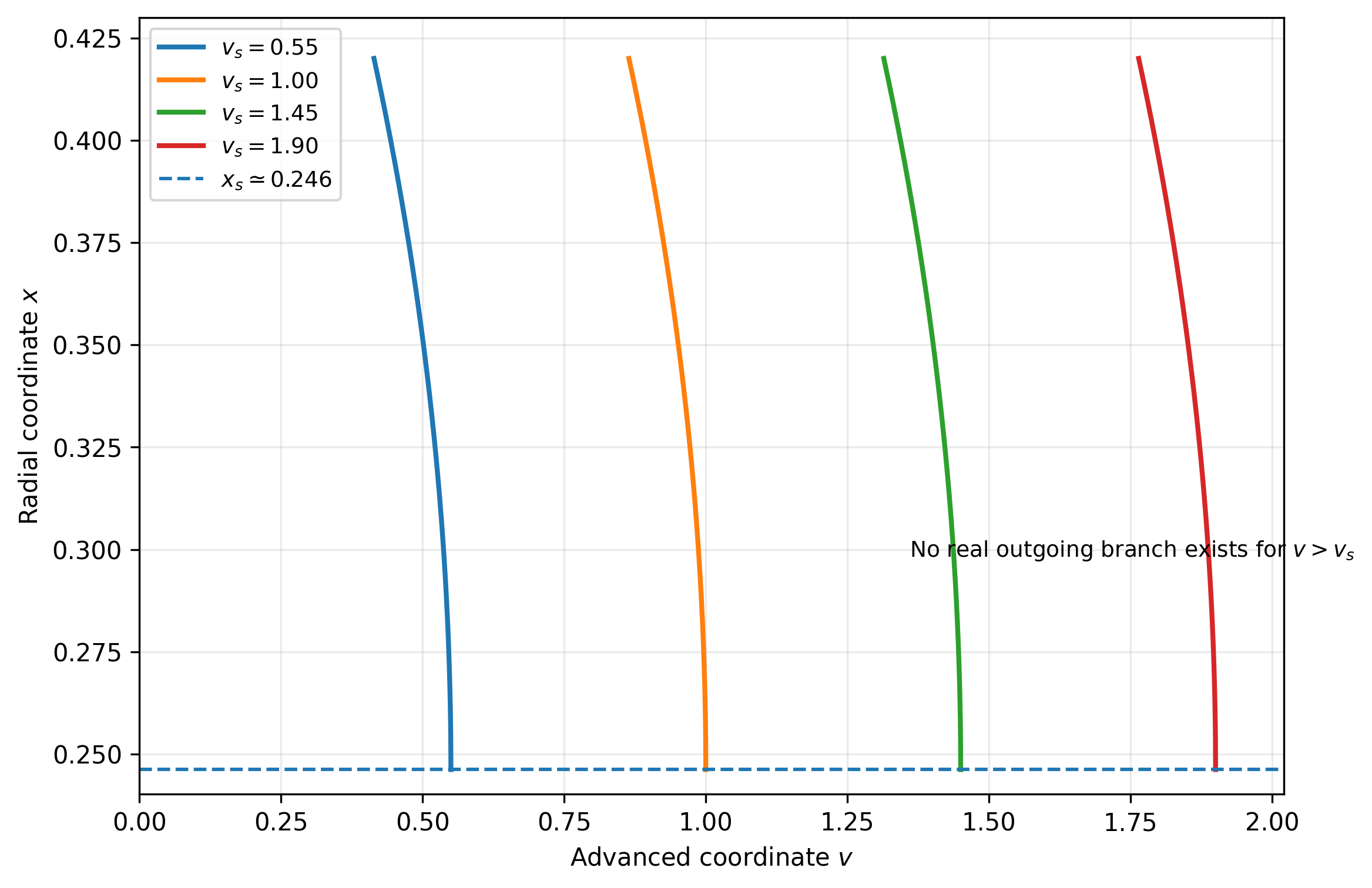}
    \caption{Local outgoing radial null curves in the negative-pole region for $\gamma=1$, $\widetilde{\omega}=-0.02$, and $q^{2}=0.10<x_s$.}
    \label{fig:3}
\end{figure}

Figure~\ref{fig:3} illustrates the local outgoing radial null curves in the negative-pole region for $\gamma=1$, $\widetilde{\omega}=-0.02$, $q^{2}=0.10<x_s$. Near $x=x_s$, the lapse has the leading-order form $f_q(x)\simeq-A_q/(x-x_s)$, where $A_q=x_s(x_s-q^{2})/D'(x_s)\simeq0.222>0$. Using the null equation, Eq.~\eqref{eq:null-eq}, and $\epsilon=x-x_s$, the leading-order equation becomes $d\epsilon/dv\simeq -A_q/(2\epsilon)$, integrating to
$
(x-x_s)^{2}=A_q(v_s-v),
$
as in Eq.~\eqref{eq:eps-neg}. Each trajectory approaches $x=x_s$ at its corresponding advanced-time value $v=v_s$; changing $v_s$ merely translates the endpoint along the advanced-time axis and does not alter the qualitative shape of the null trajectory. For $v>v_s$, the right-hand side becomes negative, so no real outgoing solution exists to the future of the singular endpoint. Hence $q^{2}<x_s$ implies the singular boundary is locally non-emitting.

This figure establishes a local non-emission property. By itself, the near-$x_s$ expansion does not provide a complete proof of global weak cosmic censorship. Nevertheless, when combined with the phase-diagram result $x_s<x_+$ and the collapse-time ordering established above, it provides strong model-level evidence that the singularity forms only after the shell has entered the black-hole region and cannot emit a future-directed outgoing radial signal.

\paragraph*{Outgoing Null Curves in the Positive-Pole Region.}\label{outgoing-positive-pole}

\begin{figure}[H]
    \centering
    \includegraphics[width=0.7\linewidth]{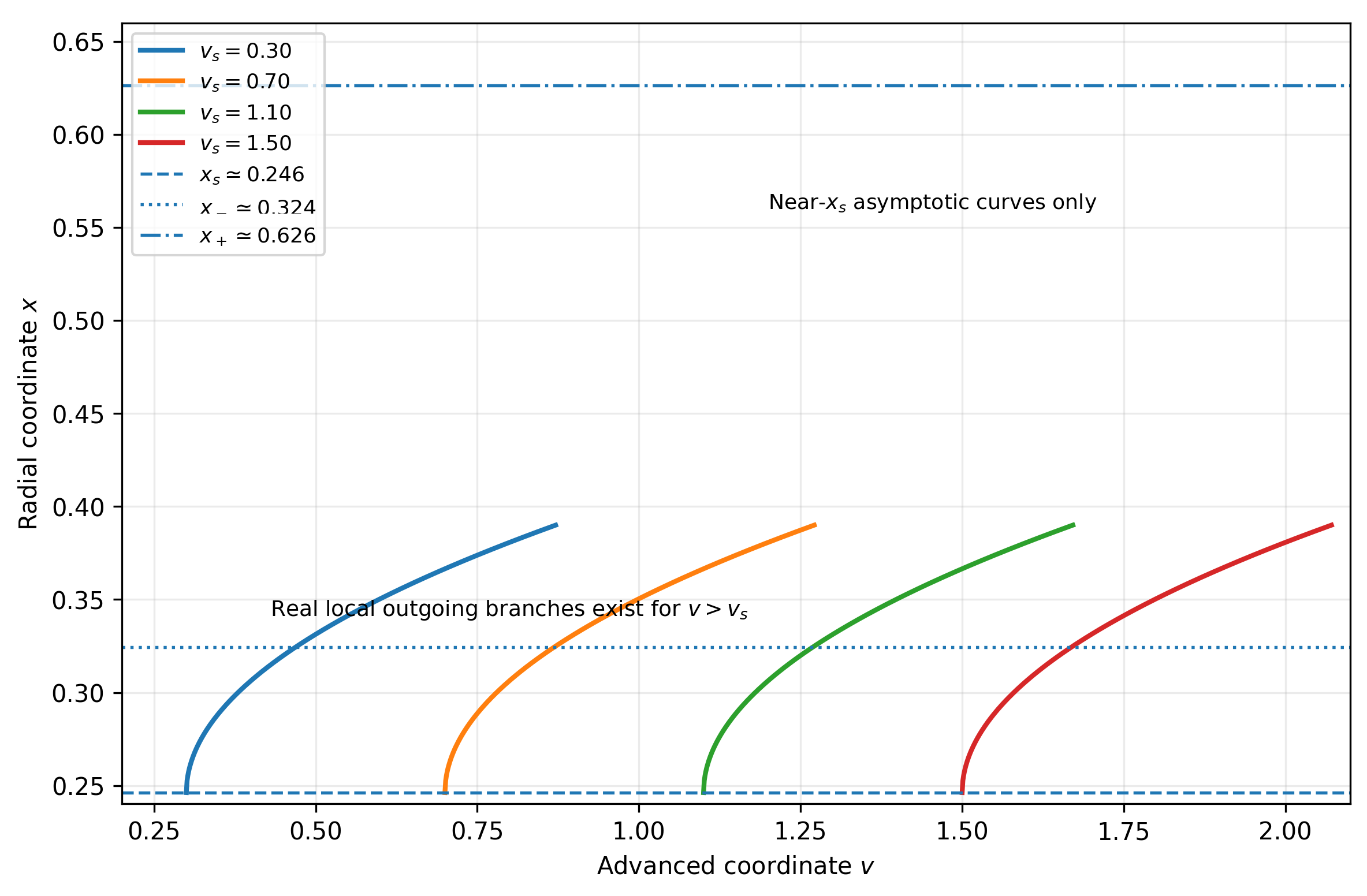}
    \caption{Local outgoing radial null curves in the positive-pole two-horizon region for $\gamma=1$, $\widetilde{\omega}=-0.02$, and $q^{2}=0.27>x_s$.}
    \label{fig:4}
\end{figure}

Figure~\ref{fig:4} shows the qualitatively different local null behavior in the positive-pole region for $\gamma=1$, $\widetilde{\omega}=-0.02$, $q^{2}=0.27>x_s\simeq0.246$. For these values, $f_q(x)\simeq C_q/(x-x_s)$, with $C_q=x_s(q^{2}-x_s)/D'(x_s)\simeq0.0362>0$. The local outgoing-null equation becomes $d\epsilon/dv\simeq C_q/(2\epsilon)$, integrating to
$(x-x_s)^{2}=C_q(v-v_s).$
In contrast to the negative-pole case, real branches exist for $v>v_s$: the curves originate locally at $x=x_s$ and increase as $x-x_s\propto\sqrt{v-v_s}$. Thus $q^{2}>x_s$ implies the singular boundary is locally emitting.

The change from $v_s-v$ in the negative-pole region to $v-v_s$ in the positive-pole region shows that reversing the sign of the lapse pole reverses the local causal behavior of the singular boundary: the negative-pole singularity is locally non-emitting, whereas the positive-pole singularity is locally emitting.

For the selected positive-pole parameters, the static exterior nevertheless contains two physical horizons, $x_-\simeq0.324$, $x_+\simeq0.626$. The existence of a locally outgoing null branch does not therefore establish that the singularity is globally naked: the singular boundary lies below the inner horizon, and an outgoing ray would still have to propagate through the complete dynamical geometry and cross the outer horizon before reaching future null infinity. Moreover, the curves shown in Figure~\ref{fig:4} follow from the leading-order near-$x_s$ asymptotic expansion and should not be extrapolated to large radii.

\paragraph*{Physical Interpretation.}\label{physical-interpretation}

Taken together, the four figures identify three distinct physical situations. First, in the negative-pole region, $0\leq q^{2}<x_s$, the lapse diverges negatively at the finite-radius curvature singularity. An outer horizon necessarily exists, the collapsing shell crosses it before reaching $x_s$, and no future-directed outgoing radial null branch emerges locally from the singular boundary. This is the parameter region supported by the horizon-shielding argument developed in the manuscript.

Second, at the tuned value $q^{2}=x_s$, the simultaneous zeros of $H_q(x)$ and $D(x)$ cancel the pole in the metric. The finite-radius curvature singularity is therefore absent, although the running gravitational coupling remains divergent. This exceptional surface must be analyzed separately from the two singular regions.

Third, in the positive-pole region, $q^{2}>x_s$, the finite-radius curvature singularity persists, and future-directed outgoing radial null branches exist locally. The global outcome depends on the horizon structure. When a physical extremal threshold exists, $x_s<q^{2}<q_H^{2}$ gives two simple horizons, $q^{2}=q_H^{2}$ gives one degenerate horizon, and $q^{2}>q_H^{2}$ gives no horizon in the physical exterior. When $x_s\geq x_c$, every positive-pole configuration is horizonless. The existence of a local outgoing branch in the positive-pole region does not by itself prove global nakedness when horizons are present.

The local non-emission and horizon-shielding conclusions of the present analysis should therefore remain restricted to $0 \leq q^2 < x_s$.

\section{Conclusion}\label{conclusion}
We have developed a charged model of scale-dependent gravitational collapse by matching a spatially flat FLRW interior to a charged, scale-dependent exterior across a timelike thin shell. The electric charge is confined to the stellar surface, thereby preserving the homogeneity and isotropy of the interior. The exterior geometry is supported by a phenomenological, Bianchi-consistent effective-source construction rather than being assumed to follow directly from the standard action-derived scale-dependent Einstein-Maxwell equations.

The Israel-Maxwell junction conditions determine the shell energy density, pressure, electric current, and equation of motion. Introducing a barotropic surface equation of state closes the dynamics and reduces the evolution to a finite set of model parameters. In the charged-dust limit, the shell material energy is conserved, and the collapse is governed by the effective potential of Eq.~\eqref{eq:Vq-dust}, subject to the unsquared junction condition and the requirement that the trajectory remain within the dynamically allowed region.

For $\tilde{\omega}<0$, the geometry is controlled by the unique positive root $x_s$ of $D(x_s)=0$. The charge separates the solutions into three physically distinct regimes. When $0 \leq q^2 < x_s$, the lapse function has a negative pole at a finite-radius curvature singularity, and the physical exterior contains one outer horizon satisfying $x_+>x_s$. At the tuned value $q^2=x_s$, the simultaneous zeros of the numerator and denominator cancel, removing the curvature pole from the metric, although the prescribed running gravitational coupling remains singular. When $q^2>x_s$, the curvature singularity persists with a positive pole. Depending on the scale-dependent parameters and the physical extremality condition $x_e>x_s$, the exterior may contain two simple horizons, one degenerate horizon, or no horizon.

The representative charged-dust-shell solution in the negative-pole region exhibits monotonic collapse without a bounce. For the parameters considered, the stellar surface crosses the outer exterior horizon before reaching the singular radius, while the interior apparent horizon subsequently intersects the surface and the trapped region expands during the final stage of collapse. This example demonstrates that the junction conditions, effective potential, exterior horizon structure, and interior trapped-surface evolution can be treated consistently within the model.

The local outgoing-null-ray analysis further distinguishes the two singular regions. For $q^2<x_s$, no real future-directed outgoing radial branch emerges from the singular boundary, whereas for $q^2>x_s$, outgoing branches exist locally. The first result, combined with horizon crossing before singularity formation along a monotonic, dynamically admissible trajectory, provides model-level evidence for horizon shielding in the negative-pole region. In the positive-pole region, local emission alone does not establish global nakedness because the rays must still be propagated through the complete matched spacetime and across any inner and outer horizons.

\section*{Statements and Declarations}
\paragraph{Funding} This research received no funding.
\paragraph{Competing interests} The authors declare no competing interests.
\paragraph{Data availability} The numerical data supporting the figures are available from the corresponding author upon reasonable request.

\end{document}